\documentclass{elsart}

\usepackage{harvard}

\usepackage{graphicx}
\usepackage{epsfig}

\usepackage{amssymb}

\def\url#1{{\ttfamily\def\/{/\discretionary{}{}{}}#1}}

\begin{document}

\begin{frontmatter}
\title{3D Structures on Relativistic Jets}


\author[hardee]{Philip E. Hardee\thanksref{ph}}, 

\thanks[ph]{E-mail: hardee@athena.astr.ua.edu}

\address[hardee]{University of Alabama, Department of Physics \& Astronomy,
 Tuscaloosa, AL  35487}

\begin{abstract}

The properties of wave-like helically twisted normal mode structures on
steady relativistic jets are summarized.  Wave speeds are a function of
the wavelength and less than the jet speed. However, normal mode
interference can lead to both stationary and superluminal phase
effects.  A maximum pressure fluctuation criterion suggested by
numerical simulations of axisymmetric relativistic jets is used to find
the maximum asymmetric jet distortions and velocity fluctuations.
Cyclic transverse velocity fluctuation can lead to variation in the
flow direction on the order of the relativistic beaming angle.
Resulting variation in the Doppler boost factor can lead to significant
brightness asymmetries as helical structures twist around the jet
beam.  Growth of these structures is reduced as the jet density, Lorentz
factor or Mach number are increased.  Maximum jet distortion is reduced
as the Lorentz factor increases and this suggests a reduction in mass
entrainment or other non-linear disruptive processes that influence the
morphological development of radio sources.

\end{abstract}

\end{frontmatter}

\vspace{-0.5cm}
\section{Introduction}
\label{intro}
\vspace{-0.4cm}

Many known or suspected relativistic jets exhibit complex
time-dependent structures, e.g., 3C~345 \cite{Zen95} \& M87
\cite{Bir95,Bir99}, that are difficult to understand in terms of simple
static jet bending and/or acceleration of components along a fixed
trajectory.  Wiggling or twisted helical structures on multiple length
scales provides evidence for wave-like ``normal mode'' structures along
these flows, and precession of the central engine provides a periodic
driver that can excite the normal modes on multiple length scales. The
amplitudes to which normal mode structures can grow has been
investigated for non-relativistic jets \cite{Hard97,Kepp99} and for
axisymmetric relativistic jets \cite{Hard98} via comparison between
theory and numerical simulations.  Here we present a prediction of the
maximum amplitudes to which asymmetric flow driven (Kelvin-Helmholtz
unstable) structures could grow on hydrodynamic relativistic jets and
the effect on observable jet structure and radio source development.
Large amplitude normal mode distortions lead to mass entrainment and
disruption of highly collimated flow \cite{Ros99,Rosen} and are an
important aspect of the morphological development of radio sources.

\vspace{-0.5cm}
\section{The Normal Modes}
\label{modes}
\vspace{-0.4cm}

A perturbation to a cylindrical jet can be considered to consist
of Fourier components of the form

\vspace{-0.3cm}
\begin{center}
$f_1(r,\phi ,z)=f_1(r)\exp [i(kz\pm n\phi -\omega t)]$
\end{center}
\vspace{-0.3cm}
 
where flow is along the z-axis, and the flow is bounded by $r=R$ in the
radial direction. In cylindrical geometry $k$ is the longitudinal
wavenumber, $n$ is an integer azimuthal wavenumber, for $ n > 0$ the
angle of the wavevector relative to the flow direction is $\theta =\tan
(n/kR)$, and $+n$ and $-n$ refer to wave propagation in the clockwise
and counterclockwise sense, respectively, when viewed outwards along
the flow direction. The azimuthal wavenumbers $n=$ 0, 1, 2, 3, etc.
correspond to pinching, helical, elliptical, triangular, etc. normal
mode distortions of the jet, respectively. For normal mode $n$ the
axial wavelength associated with a $360^{\circ }$ helical twist of a
wavefront around the jet beam is given by $\lambda _z=n\lambda _n$
where $\lambda _n=2\pi /k$.

Propagation and growth or damping of the Fourier components is
described by a dispersion relation \citeaffixed{Birk91}{see}. On the
supersonic jet, i.e., $M_{jt}\equiv u/a_{jt} >>1 $ and $M_{ex}\equiv
u/a_{ex} >> 1$ where we define the sound speeds by $a^2 \equiv \Gamma
P/\left\{\rho +[\Gamma /(\Gamma-1 )]P/c^2\right\}$ and $\Gamma$ is the
adiabatic index, each normal mode, $n$, consists of a single
``surface'' wave and multiple ``body'' wave solutions to the dispersion
relation.

In the low frequency limit all asymmetric ($n > 0$) growing normal mode
``surface'' wave solutions are given by \citeaffixed{Hard87}{e.g.}

\vspace{-0.7cm}
$$\omega /ku \approx \frac{\gamma ^2\eta _{rel}}{1+\gamma ^2\eta
_{rel}} + i\frac{\gamma \eta _{rel}^{1/2}}{1+\gamma ^2\eta _{rel}} ~,
$$
\vspace{-0.7cm}

where $\eta _{rel}\equiv (a_{ex}/a_{jt})^2$, and $\gamma$ is the
Lorentz factor. The growth rate (imaginary part) decreases and wave
speed increases as $\gamma ^2\eta_{rel} >> 1$. In the low frequency
limit all ``body'' wave solutions are given by \citeaffixed{Hard87,Hard98}{see}

\vspace{-0.7cm}
$$
kR\approx k_{nm}^{\min }R\equiv \frac{(n+2m-1/2)\pi /2}{\gamma
(M_{jt}{}^2-1)^{1/2}} ~,
$$
\vspace{-0.7cm}

where $n$ is the mode number, and $m \geq 1$ is the integer body mode
number.  In this limit $(\omega /ku)|_{real} \approx 0$ and the body
wave solutions are either purely real or damped. At higher wavenumbers,
$kR \geq k_{nm}^\ell R \gtrsim k_{nm}^{\min }R$, the body modes are
propagating and growing.

All asymmetric surface and body normal mode waves have a maximum in the
growth rate \citeaffixed{Gill65,Payn85,Hard87,Hard98}{see}, at a resonant frequency, wavelength and wave speed:

\vspace{-0.7cm}
$$
\omega R/a_{ex}\lesssim \omega _{nm}^{*}R/a_{ex}\equiv (n+2m+1/2)\pi
/2 ~,
$$
$$
\lambda \gtrsim \lambda _{nm}^{*}\equiv \frac{2\pi }{\omega
_{nm}^{*}R/u}\frac{\gamma [M_{jt}{}^2-u^2/c^2]^{1/2}}{
[M_{ex}{}^2-u^2/c^2]^{1/2}+\gamma [M_{jt}{}^2-u^2/c^2]^{1/2}}R ~, 
$$
$$
v_w\approx v_w^{*}\equiv \frac{\gamma
[M_{jt}{}^2-u^2/c^2]^{1/2}}{[M_{ex}{}^2-u^2/c^2]^{1/2}+\gamma
[M_{jt}{}^2-u^2/c^2]^{1/2}}u ~. 
$$
\vspace{-0.5cm}

In the above $m=0$ indicates the surface wave and $m\geq 1$ indicates
the body waves. At this resonance the growth rate is proportional to
$(\gamma M_{jt})^{-1}$ and growth is slowed as $\gamma M_{jt}$ increases.

At high frequencies and short wavelengths the growth rate becomes vanishingly small and the normal modes propagate like sound waves in the jet with
$$
\omega /ku \approx \frac{1 - a_{jt}/u}{1 - a_{jt}u/c^2} ~. 
$$
Thus, the normal modes propagate with a wave speed that is a function of the frequency relative to the resonant frequency, and wave speeds are always less than the underlying flow speed.

\vspace{-0.5cm}
\section{Normal Mode Structures}
\label{structure}
\vspace{-0.4cm}

Displacements of jet fluid can be written in the form

\vspace{-0.3cm}
\begin{center}
{\boldmath $\xi$}$(r,\phi ,z) = {\bf A}(r_0)e^{i{\bf \Delta }(r_0)}\xi
_r(R)\exp [i(kz_s\pm n\phi _s-\omega t)]$,
\end{center}
\vspace{-0.3cm}

where $z_s$ and $\phi _s$ are the axial and azimuthal position at the
jet surface, $r_0$ is the initial radial position, $\xi_r(R)$ is the
radial displacement at the jet surface, and the ${\bf
A}(r)e^{i{\bf \Delta }(r)}$ are simple relativistic equivalents
(Hardee 1999) of equations (A10) in \citeasnoun{Hard97}.
The velocity components associated with fluid displacements are given by

\vspace{-0.4cm}
\begin{center}
${\bf v}_1(r,\phi,z)=d${\boldmath $\xi$}$ /dt=-i(\omega -ku)${\boldmath
$\xi$}$(r,\phi,z)$ .
\end{center}
\vspace{-0.4cm}

The total velocity is ${\bf v}(r,\phi,z)={\bf u}+{\bf v}_1(r,\phi,z)$
where ${\bf u}$ is the unperturbed jet velocity along the $z$-axis.
The pressure perturbation associated with a fluid displacement can be
written in the form \cite{Hard00}

\vspace{-0.4cm}
\begin{center}
$P_1(r,\phi,z)=B(r_0)e^{i\Delta
(r_0)}\xi _r(R)\exp [i(kz_s\pm n\phi _s-\omega t)]$ .
\end{center}
\vspace{-0.3cm}

A maximal pressure fluctuation criterion, $0 < P_0 + P_1 < 2P_0$, found
from axisymmetric relativistic jet simulations \cite{Hard98}
has been used to compute the maximum expected jet distortions for
various Lorentz factors and Mach numbers \cite{Hard00}.  Figure 1 shows
the maximum normal mode jet cross section distortions for a jet of
Lorentz factor $\gamma = 10$ and with $M_{jt} \approx M_{ex} \approx
1.3$ for the helical and elliptical surface wave modes at wavelengths,
$\lambda_h^{(\imath,\jmath,*)}/R =$ 8.25, 3.85, 1.88 and
$\lambda_e^{(\imath,\jmath,*)}/R =$ 5.54, 3.49, 1.73,  where the
shortest wavelength is at resonance, and for the helical and elliptical
first body wave modes at the longest unstable wavelength, $\lambda_h^\ell/R =
3.36$ and $\lambda_e^\ell/R = 2.34$.


\begin{figure}[h!]
\vspace{7.0cm}
\includegraphics{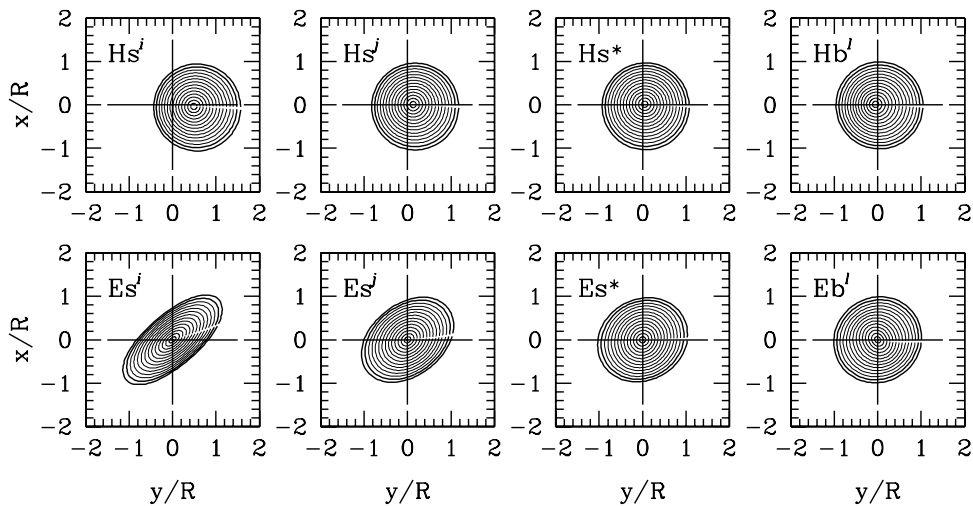}
\caption {Maximum cross section distortions for helical and elliptical
surface waves at three wavelengths ($\lambda^\imath$,
$\lambda^\jmath$, $\lambda^*$), and the first body wave at the longest
unstable wavelength, $\lambda^\ell$.}
\label{fig_1}
\end{figure}

Jet distortion is a function of the wavelength and scales relative to
the resonant wavelength of the appropriate normal mode and not, for
example, to the absolute value of the wavelength.  High pressure
regions are coincident with regions where the displacement surfaces are
closer together. The cross sections rotate counterclockwise as the
waves propagate outwards along the jet and, in general, high pressure
regions precede a trailing low pressure region where the maximum
outwards surface displacement occurs. For a given pressure fluctuation,
for example at the resonant wavelength, the accompanying jet distortion
decreases as the Lorentz factor increases.  Thus, in general, the cross
sections of jets with large Lorentz factors should be less distorted
than those with small Lorentz factors.  Still at long wavelengths the
helical and elliptical surface modes can lead to significant jet
distortion.  Other results suggest that most observable (large scale)
structure will develop at wavelengths equal to or greater than the
resonant wavelengths of the normal modes.  Nevertheless, tightly
wrapped helical structures are possible on hot relativistic jets with
large Lorentz factors because some resonant wavelengths can be less
than the jet radius.

The velocity and pressure fluctuations  that accompany the distortions
can lead to interesting consequences for Doppler boosted emission
features associated with asymmetric structures.  We can investigate the
possible effects on emission features by constructing an ``apparent''
emissivity, $P^2D^2$, around the displacement surfaces shown in Figure
1 where $P$ is the pressure and $D=\{\gamma [1-(v/c)\cos \theta
]\}^{-1}$ is the Doppler boost factor. Let us assume that the initial
unperturbed flow angle with respect to the the line of sight is $\theta
_0=1/\gamma _0=0.1$ rad, the beaming angle.

Figure 2 gives the values of $P^2D^2$, $\gamma$, $\theta$ and $D$ for
the surface modes at the longest wavelength, $\lambda^\imath$, shown in
Figure 1. Values are plotted around displacement surfaces with
unperturbed radial location of $r_0/R=1/7,2/7,...,7/7$. The azimuthal
angle $\phi =0$ lies along the $+y$-axis in Figure 1 and increases in
the clockwise sense.


\begin{figure}[h!]
\vspace{6.0cm}
\includegraphics{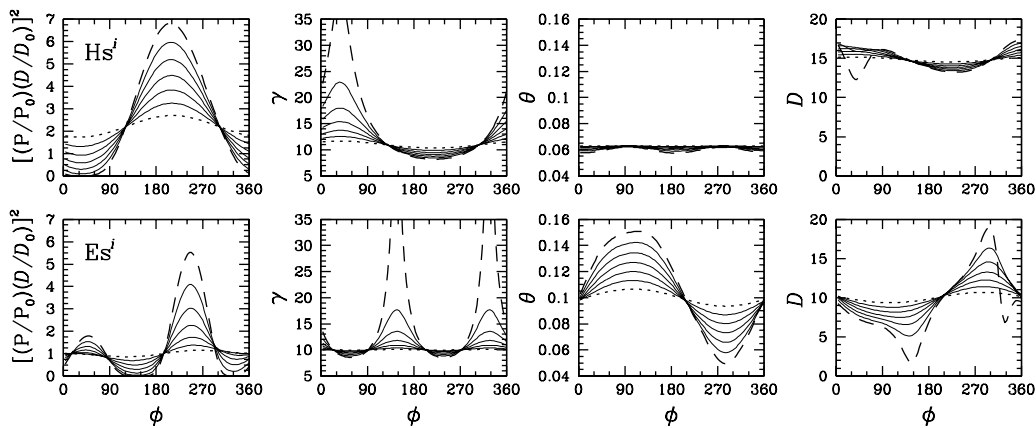}
\caption {Emissivity, Lorentz factor, flow angle and Doppler boost
factor around the displacement contours for the surface wave at the
longest wavelength, $\lambda^\imath$, shown in Figure 1.  The dashed
(dotted) lines correspond to the surface (center) contours in the jet
cross sections, respectively.}
\label{fig_2}
\end{figure}

For the surface waves the highest pressures and highest
``emissivities'' occur at the jet surface and for higher order surface
modes the distortion amplitude and pressure fluctuation decrease
rapidly away from the jet surface.  Variation between maximum and
minimum apparent emissivity as emission enhanced regions twist around
the jet beam, is about a factor of six for the helical mode (not shown)
and about a factor of three for the higher order elliptical surface
mode, e.g., the difference in emissivity for the two elliptical filaments
in Figure 2.

Figure 3 gives the values of $P^2D^2$, $\gamma$, $\theta$ and $D$
around displacement surfaces for the first body modes at the longest
unstable wavelength, $\lambda^\ell$. For the body modes the highest
pressures and emissivities occur in the jet interior.  The variations
that occur at the resonant wavelength are similar to those shown in the
Figure 3 for the longest unstable wavelength.  Surface mode variations
at the resonant wavelength are also similar to those shown in Figure 3
although the radial structure is different.


\begin{figure}[h!]
\vspace{6.0cm}
\includegraphics{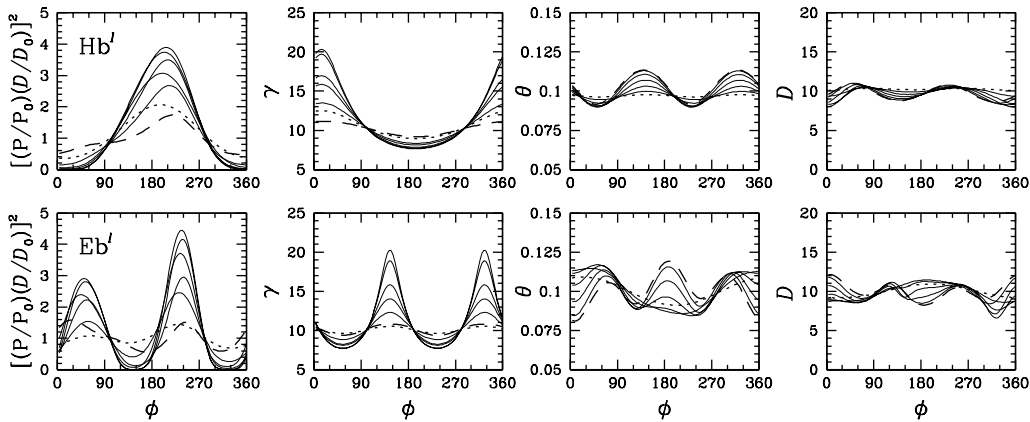}
\caption {Like Figure 2 but for the body wave at the longest unstable
wavelength shown in Figure 1.}
\label{fig_3}
\end{figure}

The asymmetric normal modes lead to high pressure regions helically
twisted around the jet beam.  High pressure regions may be confined
close to the jet surface, penetrate deeply into the jet interior or be
confined to the jet interior and range from relatively thin sheets and
ribbon like to thicker and tube like depending on the mode and
wavelength.  In general, pressure and emissivity maxima are in phase
azimuthally in the jet interior at long wavelengths.  At shorter
wavelengths the jet center leads the surface and the high pressure
region adopts a trailing spiral configuration between jet center and
surface.  Although axial velocity fluctuation is a small fraction of
the relativistic jet speed, significant variation in the Lorentz factor
occurs on highly relativistic jets.  Even at high
Lorentz factor the velocity fluctuations allow for significant angular
variation in the flow direction relative to the beaming angle.

\vspace{-0.5cm}
\section{Implications}
\label{finale}
\vspace{-0.4cm}

Jet precession provides a periodic source for the generation of
multiple normal mode structures. The interference between normal mode
structures of different wavelengths and moving at different wave speeds
allows for phase effects producing superluminal and relatively
stationary features.  Stationary features are created at specific
distances from the origin where normal modes of different wavelengths
are in phase \cite{Xu00}.  Superluminal like motions have not yet been
observed in numerical simulations.  Non-relativistic numerical
simulations have shown resonant pattern motions on the order of the
resonant wave speed with motions of localized features within about
$\pm15\%$ of the average \cite{Hard99,BirHard}.

High Mach number, jet density and Lorentz factor reduce the growth rate
of normal mode structures and high Lorentz factor reduces the maximum
amplitude of jet distortions.  Nevertheless, significant variation in
the Lorentz factor and significant angular variation in the flow
direction relative to the beaming angle can occur on relativistic
jets.  Ultimately, if the jet is sufficiently long, large amplitudes at
long wavelengths (relative to resonance) can develop and lead to mass
entrainment, disruption of highly collimated flow \cite{Ros99,Rosen},
and influence the morphological development of radio sources over their
lifetime \cite{Eilek}.

The jet in M87 provides an illustration of twisted structures on
multiple scales that indicate normal mode structures.  Radio
observations \cite{Owen89} provide strong evidence for a single bright
emission filament twisted around the jet beam inside knot D and a
fainter twisted filament pair between knots D and F.  These features
indicate the presence of multiple body and surface modes.  Similar
structures are evident in VLBI images at parsec scales \cite{Reid89}.
The average motion of well defined structures along the M87 jet is
subluminal, $\approx 0.3c$, and should be taken as indicative of a wave
speed and not the speed of the underlying flow.  However, motions less
than 0.03c and up to about 6c are also observed
\cite{Reid89,Bir95,Jun95,Bir99}.  Interestingly, the highest observed
motions occur for individual bright knots often located in regions with
complex structure.  At HST-1 \cite{Bir99} rapidly moving components
appear to emerge from a subluminally moving component accompanied by
brightness variations up to $\pm 20\%$.  This type of behavior is
consistent with interference between different wave modes and line of
sight phase effects.  In the knot A, B and C complex, the observed
bending and oscillation in jet width may be identified with large
amplitude helical twisting and elliptical distortion associated with
the surface modes. This rapidly growing distortion then leads to mass
entrainment and disruption of the highly collimated jet flow.


\end{document}